# Studying the electronic and phononic structure of new graphene allotrope: Penta-graphane

H. Einollahzadeh[1], R. S. Dariani[1,*], and S. M. Fazeli[2]

[1]Department of Physics, Alzahra University, Tehran, 1993893973, Iran

[2]Department of Physics, University of Qom, Qom, 3716146611, Iran

*Corresponding author. Tel: +98 21 85692646; fax: +98 21 88613935

Email address: dariani@alzahra.ac.ir

**Abstract**

In this paper, we introduce a new structure, similar to the graphane, which we call it penta-graphane. This structure is obtained by adding hydrogen to penta-graphene $sp^2$ bond. We study the electronic and phononic structure of penta-graphane. We firstly use DFT with the generalized gradient approximation RPBE to compute the band structure. Then one–shot GW ($G_0W_0$) approach for estimating accurate band gap is applied. The quasi-band gap of penta-graphane is obtained 5.78 eV, which is nearly equal to band gap of diamond. Therefore, this new structure is a good insulator and can be alternative for electrical insulator. We also investigate the stability of penta-graphane by computing the phonon structure. Finally, by the phonon density of state, we calculate its specific heat. The high specific heat of penta-graphane releases its capability for storing and transferring energy.

**Introduction**

Carbon is interesting for scientists because of its benign environmental properties, and it is one of the most abundant elements in the nature. Carbon can bond with itself in at least three different ways, which gives us three different materials. In diamond each carbon atom is strongly bonded to four adjacent via $sp^3$ hybridization, in graphite each carbon atom is bonded three neighbors via sp$^2$ hybridization. In carbon nanotube, carbon atoms have a mixture of sp$^3$ and sp$^2$ hybridizations [1-2].

In 2004, a group at Manchester university discovered graphene; a single-atom- thick layer of carbon from bulk graphite, by a very simple and effective method, namely "scotch tape method" [3].

Graphene is the first 2D crystal ever known to us. In graphene, atoms are tightly packed in a honeycomb crystal lattice. The band gap of graphene is zero, which shows that graphene is a semimetal. Graphene is a unique super material and is the thinnest, lightest and strongest material ever known. It is transparent and conducts electricity much better than carbon [4]. Nowadays hundreds of laboratories around the world deal with different aspects of graphene. An interesting area in these researches is investigating new graphene allotropes, since 2D dimension nanostructures are ideal for modern electronics [5-8].

In 2012, new phase of carbon was reported. T12-carbon is a tetragonal allotrope of carbon, with 12 atoms in unit cell; it is stable and has the physical properties of diamond phase [9]. In early 2015, a 2D graphene allotrope which composed of pentagon structure is proposed. This structure can be exfoliated from T12-carbon [10]. The band gap of penta-graphene is obtained about 4.1-4.3 eV by $G_0W_0$ approximation [11]. Penta-graphene is metastable and has got unique properties, which may leads to broad applications in nanotechnology.

Penta-graphene has sp$^2$ and sp$^3$ bonded carbon atoms. The sp$^2$ bonds have got an extra electron, so we can hydrogenate penta-graphene. As we know the fully hydrogenate graphene is called as graphane [12-13], therefore, we call this new structure as penta-graphane in this article. Here, we use DFT and $G_0W_0$ approximation for determining the electronic structure and properties of penta-graphane.

Also, we study the phonon dispersion and the phonon density of states for penta-graphane by using DFT-GGA (RPBE) calculation in the framework of a pseudopotential approach. Then we compare the heat capacity of penta-graphene and penta-graphane.

**Computational method**

All calculations in this work are carried out within the *ab initio* ABINIT package [14]. Density functional theory (DFT) is a valuable method for description of many body systems through the electron density. The DFT is the theory due to Kohn-Sham (KS). The main goal of (KS-DFT) theory is to approximate exchange-correlation functional (xc) or corresponding potential, more accurately. The generalized gradient approximate (GGA) is an approach for estimating xc, in GGA exchange correlation energy depends on both the electron density and the gradient of the electron density. The DFT is restricted to the ground state only; therefore DFT often underestimates the band gap of semiconductors and insulators [15-16].

The GW is a good method for correcting the band structure. This method relies on a perturbative treatment starting from the DFT. In GW, the self-energy is approximated as the product of Green function G and the screened Coulomb interaction W within the random phase approximation [17]. The self-consistent GW computations are cumbersome, so the most commonly approximate is applying a first- order perturbative approach called one-shot GW ($G_0W_0$). In the $G_0W_0$, the quasiparticle wave functions can be approximated by the Kohn-Sham orbital, thus the quasiparticle energy becomes:

$$E_i = \varepsilon_i + Z_i \langle \Phi_i \mid \Sigma(\varepsilon_i) - V_{xc} \mid \Phi_i \rangle, \tag{1}$$

where $\Sigma = G_0W_0$ and $Z_i$ is the renormalization factor [18-19].

The phonon frequency $\omega$, as a function of the phonon wave vector **q**, is the solution of the following secular equation for the force constant $C_{s(\alpha),t(\beta)}$ :

$$det \left| \frac{1}{\sqrt{M_s M_t}} C_{s(\alpha),t(\beta)}(\boldsymbol{q}) - \omega^2(\boldsymbol{q}) \right| = 0, \tag{2}$$

Here, $M_t$ and $M_s$ are the atomic masses. The dynamical matrix is denoted by

$$C_{s(\alpha),t(\beta)}(\mathbf{q}) = \frac{\partial^2 E}{\partial u^*_{s(\alpha)}(\mathbf{q}) \partial u_{t(\beta)}(\mathbf{q})}, \tag{3}$$

where $u^{\alpha(\beta)}_{s(t)}$ is the displacement of atom s(t) in $\alpha(\beta)$ direction. A discrete Fourier transport leads to the real space force constant. We can use the dynamical matrix in real space or reciprocal space for phonon calculation. Useful details on investigating the response to atomic displacement are in ref [20].

**Results and discussion:**

**Electronic structure**

By hydrogenating penta-graphene, 2D crystalline phase can be obtained theoretically, which because of similarity to graphane, we call it penta-graphane. As we mentioned, carbon atoms in penta-graphene has two $sp^2$ and $sp^3$ hybridizations, and by adding hydrogen to $sp^2$ carbon bonds, penta-graphane has only $sp^3$ hybridization. The symmetry of penta-graphene is p-$42_1$m, which can be specified by tetragonal lattice, thus hydrogenation of penta-graphene does not change its symmetry. The unit cell of penta-graphane consists of six carbons and four hydrogens, the ratio of C/H is 3/2, which distributed symmetrically in penta-graphane. This structure is figured out in Fig.1. In Fig. 1(a) the unit cell of the penta-graphane is shown with dashed line. Hydrogens labeled by "d" points are located at the bottom of the C2 atoms, and hydrogens labeled by "u" points are located at the top of the C2 atoms.

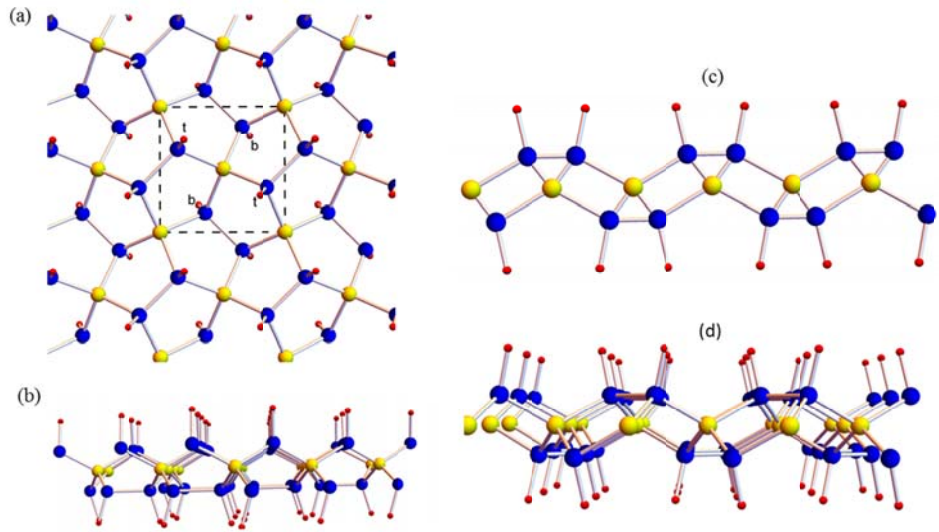

Fig.1. (a) Top view and (b) side view (c) front view (d) front view(three dimensional) of a the atomic structure of penta-graphane. Yellow, blue and red sphere represent C1, C2 and hydrogen atoms; respectively.

We compute the optimum structure of penta-graphane by ABINIT, the optimized lattice constants are found to be a=b=3.60 Å, which are 1% smaller than penta-graphene. Penta-graphane has two types of carbon, C1 has $sp^3$ bond with three carbons and C2 has $sp^3$ bond with two carbons and one hydrogen. The bond length of (C1-C2) is 1.55 Å, (C2-C2) is 1.53 Å and C2-H is 1.13 Å. The bond angle $\theta_{C2-C1-C2}$ is 124.2°, the thickness of this structure becomes 1.45 Å.

Here we compute the band structure of penta-graphane using DFT-GGA by the "Revised Perdew–Burke–Ernzerhof" type (RPBE) functional [21] (based on the Kohn-Sham eigenvalues). We have a two dimensional unit cell, since we have to use the supercell technique in ABINIT, the vacuum along the $z$ direction is taken to be 10Å, which is sufficiently large for converging the total energy. The ABINIT wave functions are expanded in plane waves, the plane wave basis set is used with a cutoff energy of 50 Hartree, which total energy converges to 1 meV/atom. The Brillouin zone sampling k point mesh parameter is 18 18 1 which is based on Monkhorst-pack grid [22].

By using DFT- GGA (RPBE), we conclude the penta-graphane has indirect band gap, which its valance band maximum (VBM) is in the Γ point, and conduction band minimum (CBM) is in (0.389 0 0) close to the M point in Γ-X path. Here, we apply Martins-Trouiller pseudopotential [23], 2s electrons are applied in potential. By considering penta-graphane unit cell, we have 28 electrons, so penta-graphane have 14 filled band and the band gap is located between $14^{th}$ and $15^{th}$ levels. The band gap of penta-graphane becomes 4.59eV by DFT- GGA (RPBE) approximation. The difference between indirect and direct gaps is 0.5 eV. The band structure of penta-graphane around Fermi level by DFT-GGA (RPBE), along symmetry line Γ-X-M-Γ (with $\Gamma = (0,0,0), X = \left(0,\frac{1}{2},0\right)$ and M= $(\frac{1}{2},\frac{1}{2},0))$ is shown in Fig.2 (a).

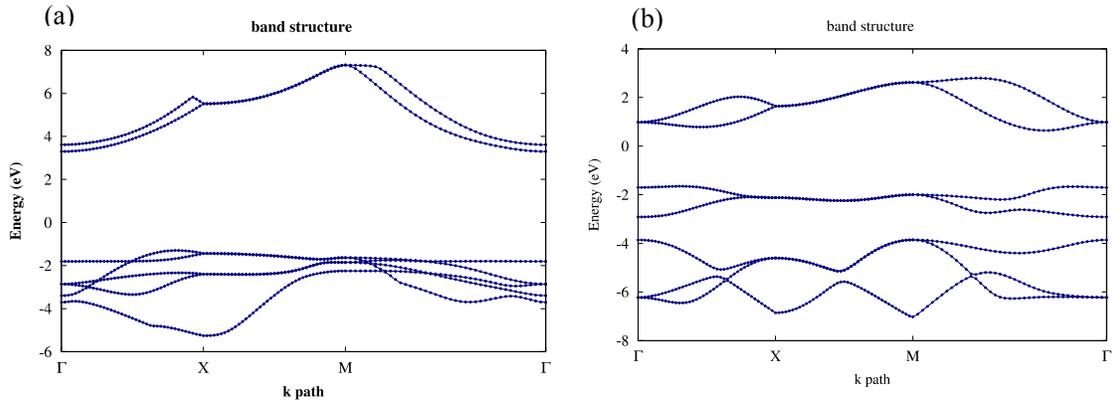

Fig. 2. (a)The band structure of penta-graphane and (b) penta-graphene [11] calculated by using DFT-GGA (RPBE) around Fermi energy.

The difference between the minimum and the maximum of the highest valance band in penta-graphane and penta-graphene is 0.37 eV and 0.58 eV, which leads to a higher total density of state (DOS) near Fermi level in penta-graphane (Fig.3(a)). Therefore by p doping, we have very high DOS in Fermi level which leads to the strong electron-phonon coupling. Because of the strongly relation between DOS at Fermi level and superconducting transition temperature, we could expect the high superconducting transition temperature [24]. The analysis of partial DOS indicates that the total DOS near VBM and CBM is mostly related to S and P states, and the total DOS below Fermi level is mainly composed of H(s) and C2(p) (Fig.3 (b) and (c)).

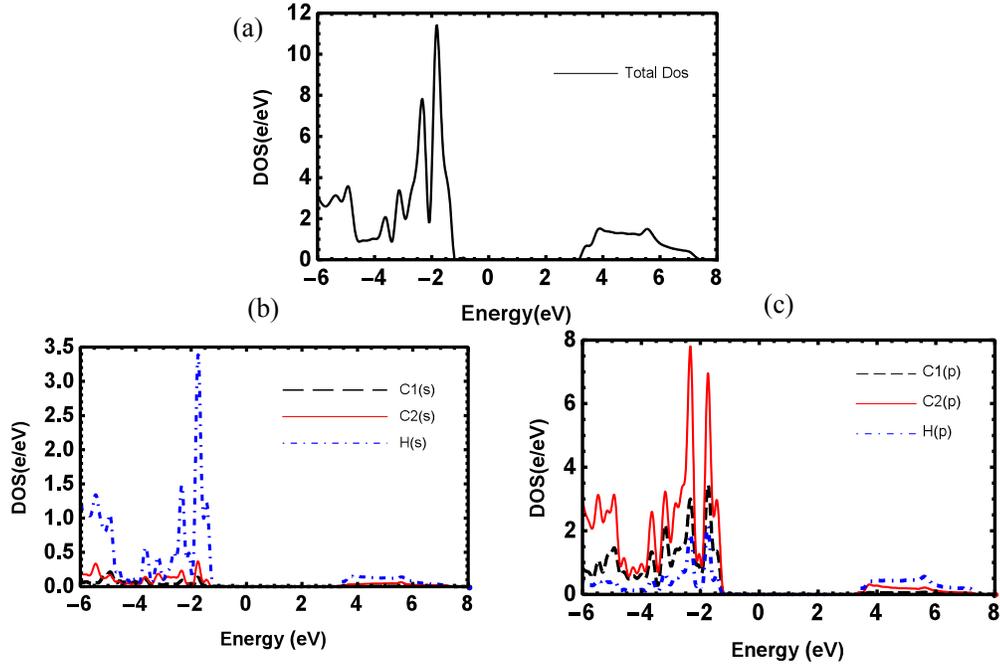

Fig. 3. (a) Total density of states , (b) Partial density of s- states and (c) p- states of penta-graphane ,(balck, blue and red corresponds to C1,C2 and hydrogen).

For a better approximation, we use $G_0W_0$. $G_0W_0$ corrections are obtained using 18 18 1 $k$ points in Brillouin zone and 20 Hartree cut-off energy. The computations converge at $n$=50 bands for computing W and self-energy. Using interpolation, we have plotted full band structure of penta-graphane in Fig. 4. This plot shows that penta-graphane has indirect band gap and the band gap of penta-graphane in $G_0W_0$-GGA is 5.78 eV. Therefore penta-graphane has a band gap sufficiently to be an electric insulator. Note that the band gap of penta-graphene in $G_0W_0$-GGA is 4.14eV [11]. The reason for this band gap opening, when penta-graphene is hydrogenated and forms penta-graphane is that $sp^3$ bond are saturated and we have not any free $\pi$ orbital. The fluctuation of valance band is due to the perturbative solving of GW that occurs when bands are very close together.

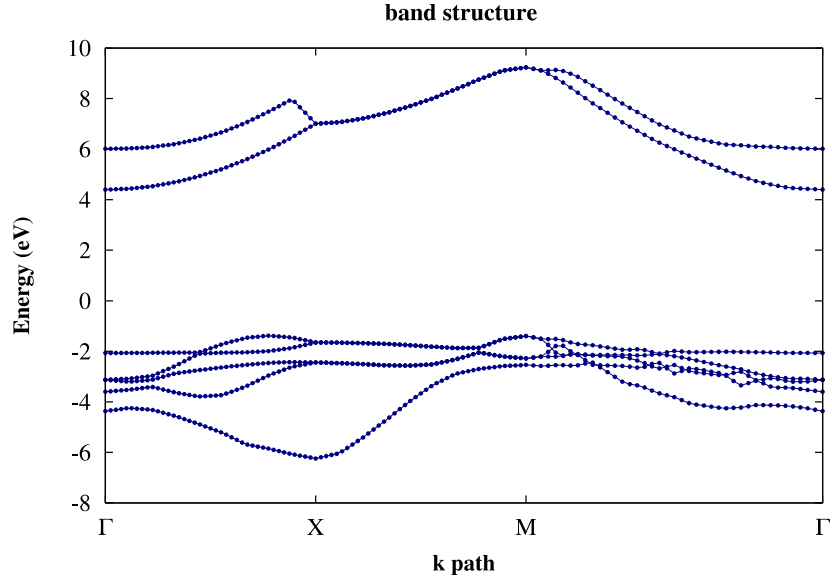

Fig. 4. The band structure of penta-graphane by using GW.

Comparing the band gap of carbon allotropes, penta-graphane has got the largest band gap, a bit larger than diamond and graphane $G_0W_0$ (5.56eV and 5.66 eV [25-26] respectively). Therefore, penta-graphane has a good electrical resistance due to this band gap, and since penta-graphane is a covalent solid, it also has a good thermal conductivity.

**Phonon structure:**

We study the stability of the two dimensional penta-graphane by calculating the phonon dispersion curves reported in Fig. 5. Phonon structure is obtained with ABINIT code. The vacuum between layers is the same as the electronic calculations, 10 Å. The dynamical properties are calculated with the density-functional perturbation theory [20]. We use Fourier based interpolation of the interatomic force constants.

The 10 atoms per primitive unit cell give us 30 vibration modes which have three acoustic modes with zero frequency at Γ point, and 27 optic modes. The absent of any imaginary frequency signifies that the penta-gtaphane is stable. The high frequency modes are higher in frequency in comparison with penta-graphene [10], and this high frequency shows that penta-graphane is a covalent solid.

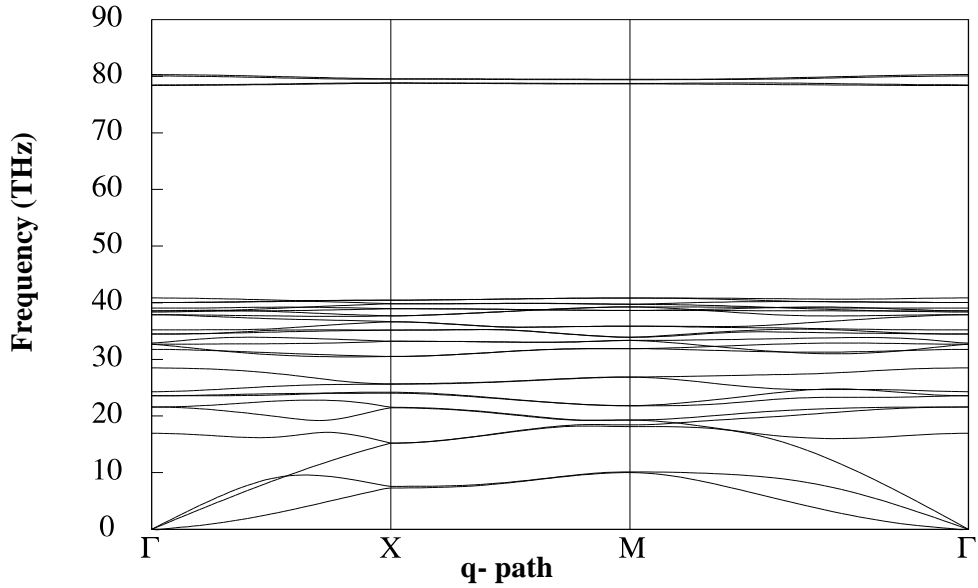

Fig. 5. Phonon dispersion of penta-graphane.

The total phonon density of states (PDOS) and partial PDOS are shown in Fig 6. The low frequency band (acoustics frequency) (0-18 THz) corresponds to C1 and C2 vibrations. Hydrogen has lighter mass, so the high frequency band is mainly H-like. The frequency band between (20-30 THz) is C2 –like and between 30-40, C1, C2 and H have nearly the same vibrations. The analysis of partial density of state shows that we have C-H ($sp^3$) stretching modes around 80 THz (2600 cm$^{-1}$).

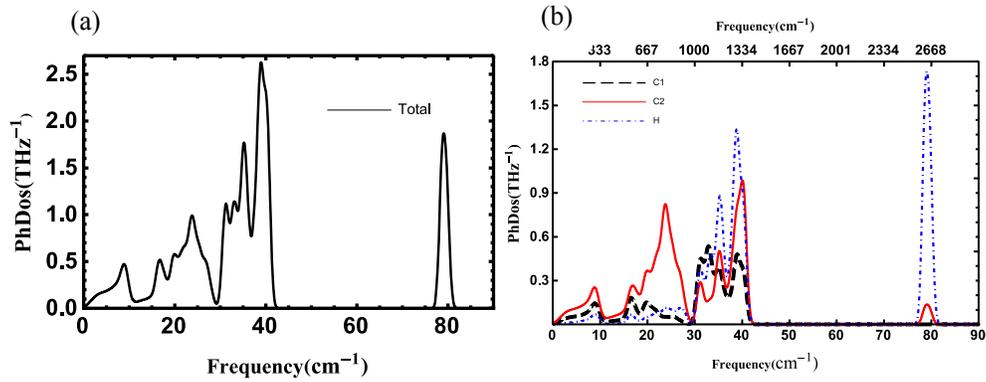

Fig. 6. (a) Total phonon density of state (b) partial contribution from C1,C2 and H (PDOS).

From the PDOS we can calculate the specific heat at constant volume by taking the second derivation of the total energy:

$$C_V(T) = k_B \int_0^{\omega_{max}} \left(\frac{\hbar\omega}{k_B T}\right)^2 \frac{\exp\left(\frac{\hbar\omega}{k_B T}\right)}{\left(\exp\left(\frac{\hbar\omega}{k_B T}\right)-1\right)^2} g(\omega)d\omega. \tag{4}$$

We compare the specific heat at constant volume of penta-graphene and penta-graphane in Fig 7.

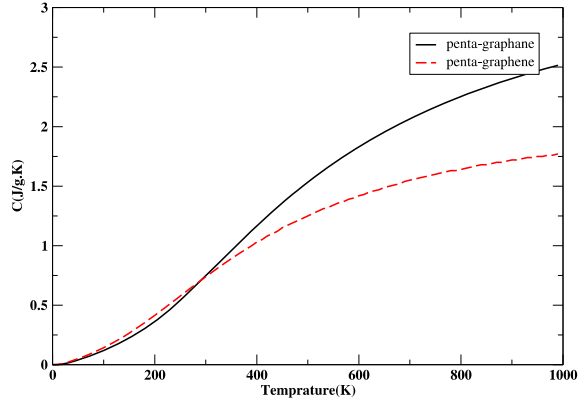

Fig. 7. The specific heat at constant volume of penta-graphene and penta-graphane

The reason for the higher specific heat in comparison with penta-graphene is the existense of hydrogen bondings .

**Conclusions**

In this paper we have proposed a new allotrope of graphene, pentagraphane. The band structure of penta-graphane has been computed by DFT and $G_0W_0$ which we have shown in Fig. 2 and Fig. 3, respectively. We have found that penta-graphane is a good insulator with 5.78 eV band gap. Penta-graphane has diamond-like structure with $sp^3$ hybridization. It has a good thermal conductivity due to its band gap. Penta-graphane has high total density of state near Fermi level, therefore p doping leads to strongly DOS in Fermi level. By this high electronic density of state at Fermi level and its phonon spectrum, we would expect high superconducting transition temperature. The phonon dispersion of penta-graphane has been shown in Fig. 5. The absent of any imaginary frequency signifies the stability of the penta-graphane. By the computed stability and electronic structure of this new two-dimensional carbon allotrope, we can expect to have potentially interesting application. The specific heat of the penta-graphane is calculated by the phonon density of state, which releases high capability for storing and transferring energy of penta-graphane.